\documentstyle[aps,prl,epsfig,float]{revtex}
\draft
\begin{document}
\twocolumn[\hsize\textwidth\columnwidth\hsize\csname
@twocolumnfalse\endcsname
\title{ Polar Smectic Films}
\author{Isabelle Kraus \cite{perm}  and Robert  B.\ Meyer}
\address{The  Martin Fisher   School  of Physics,
Brandeis University, Waltham, MA 02254-9110}

\date{\today}
\maketitle

\begin{abstract}

We   report on a new  experimental  procedure for forming and
studying polar smectic liquid  crystal films.  A free  standing
smectic film is put in  contact with a  liquid drop, so that  the
film has  one liquid crystal/air interface and  one liquid
crystal/liquid  interface.  This polar environment results  in
changes in the  textures observed in the film,  including a boojum
texture and   a previously unobserved spiral texture in which the
winding direction of  the  spiral reverses at  a finite radius
from its center. Some aspects of these textures are explained by
the presence of a $K_{sb}$-term in the bulk elastic free energy
that favors a combination of splay and bend deformations.

\end{abstract}

\vskip2pc]

When considering soft-condensed-matter composed of organic
molecules, interfaces have a strong influence on the  molecular
arrangement within the sample. This is particularly true for thin
films (from a few to a few hundred nanometers thick) where the
interface contributions are not overwhelmed by the bulk
contributions.    Thus, thin ordered  films   are a good probe  to
investigate the consequences for the molecular arrangement when an
asymmetrical environment is imposed on   the sample. A polar
geometry can be  obtained   when the  film is  placed between two
different isotropic media: liquid at one  interface and air at the
other.   Illustrations  include bi-phasic systems, {\it  e.g.},
pre-wetting films  induced on top of a drop of mesogenic liquid
\cite{drop}, as well as bi-chemical systems where the liquid
medium is   no longer a melted  form  of the  ordered wetting
phase  but is of a different chemical nature. Focusing on
bi-chemical systems, recent work involves nematic films spread on
glycerin \cite{LAV}, as well as Langmuir multi-layer \cite{004}
and mono-layer \cite{LM}  films spread on water. Here, the
molecular head-tail asymmetry combined with a 2-dimensional
confinement (a circular domain within the film) is the source of
rich and complex molecular patterns, of which one called a boojum
has generated much interest in recent years \cite{006}.

In this Letter, we  report a new experimental  approach to achieve
an asymmetric environment    using  free  standing  films    of
smectic liquid-crystal   (LC)  material, deposited  onto    a
liquid drop  by contact. This method makes use of much simpler
techniques and smaller scale apparatus than the ones  commonly
used to study Langmuir films, for  example. It also  provides
researchers with  a  broad range  of experimental possibilities
such as wetting the  film with any kind of liquid   as long  as it
does  not dissolve  the liquid  crystal, and choosing the  initial
thickness  of the film.  Motivated by  the rich range    of
pattern formation  in  free   standing smectic films,  we
developed  this technique especially to  study the changes in
pattern formation  when   a free   standing   film  was put   into
a  polar environment. With that in mind, we examined polar films
of the tilted chiral  smectic~C* (Sm~C*) phase of several  liquid
crystals on a variety of liquid substrates. We paid particular
attention to single 2D confined domains   in which we observed
significant changes  in the textures involving a topological point
defect, including a boojum pattern and, most remarkably, a spiral
pattern reversing its direction of winding at a particular radius.

\begin{figure}[h]
\centerline{\epsfig{file=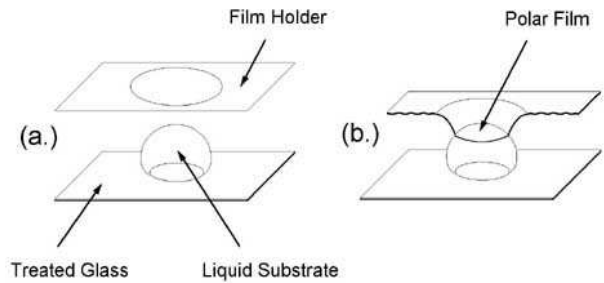,height=4cm}} \caption{
Experimental set-up: (a) Before contact, a free standing film  is
spread on the holder and a drop of liquid is deposited on the
slide.  (b) After contact, the polar film is created on top of the
drop.} \label{1}
\end{figure}

%\section{System}
Our   samples  are   obtained starting    from  a free standing
thermotropic smectic film, formed by spreading the smectic across
a circular  frame (a 6 mm diameter hole in a thin metal plate).
The film consists of a stack of molecular layers, each about 30
\AA\ thick, oriented parallel to the two free surfaces in contact
with air, defining a film of homogeneous thickness.  Using a
micrometer screw, the  film is slowly  lowered toward the top  of
a liquid drop. The drop (of,  say, water) is held in  place in a
small hydrophilic  circle on  an  otherwise hydrophobically
treated glass substrate. When the  contact is established,  a
polar film with one LC/air interface and one LC/liquid interface
is formed on top of the drop.  The resulting sample consists of
the polar film itself  as shown in Fig.\ \ref{1} and of the
``remaining'' free  standing film stretched between  the film
holder and the liquid  medium.

On the drop, a circular meniscus of extra LC material forms at the
junction  of the  free standing film  with  the  polar film.
Depending on the  relative surface tensions   of the three
surfaces meeting   at this meniscus, the top   of the drop  is
flatter than initially, still tending to be somewhat curved. The
diameter of the polar film can be  varied by raising or lowering
the film holder position. The original  free standing film can be
made in thicknesses ranging from thousands down to two layers
\cite{010}.

%section{Samples we examined}

We made  samples of hundreds of layers as well as of about 30
layers. Extremely thin samples  are difficult  to deposit  by this
method.   We  examined films  on distilled water, ethylene glycol,
and aqueous solutions of Poly-Vinyl-Alcohol (PVA) and Sodium
Dodecyl Sulfate (SDS). We ran the experiment at  $20 ^\circ$
Celsius in order to  minimize  the evaporation of the wetting
liquid. Most of our  studies were made on films  of the room
temperature  ferroelectric Sm~C*  mixture CS1015 from Chisso
\cite{chem}.  Sm~C*  is a layered liquid crystal phase composed of
chiral (non-reflection symmetric) molecules. The mean molecular
long axis, given  by the director $\hat{n}$, chosen to point
``upwards'' from the film, is tilted relative to the layer normal
through an equilibrium angle $\psi$ (26 degrees for CS1015). The
projection  of $\hat{n}$ onto the plane of the film defines the
unit vector field, the $c$-director $\hat{c}(x,y)$. The chirality
of Sm~C* results in a spontaneous rotation of $\hat{c}$ as a
function of $z$, the direction normal to  the plane of the film.
Given the helix pitch of the Sm~C* phase  for  CS1015 of  3
$\mu$m,  and our  maximum film thickness of   a  few hundred nm,
we neglect this effect and we consider $\hat{c}(x,y)$ uniform
along $z$ \cite{008}.

\begin{figure}
\centerline{\epsfig{file=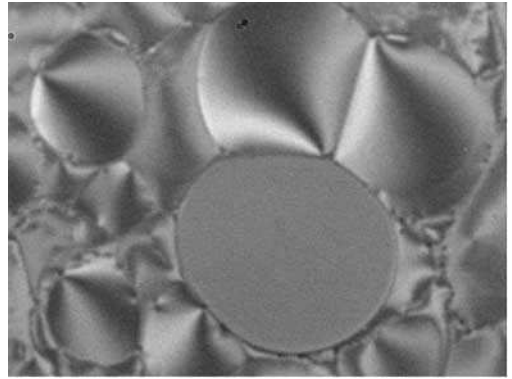,height=5cm}}\vspace{0.25cm}
\caption{ Polar smectic~C* film observed between slightly
uncrossed (vertical and horizontal) polarizers. The texture shown
in the roughly circular domains, a boojum, is the most commonly
observed one. The grey area is a thin (tens of layers) Smectic A
domain. Image width: 156 $\mu$m.} \label{boojum}
\end{figure}

\begin{figure}
\centerline{\epsfig{file=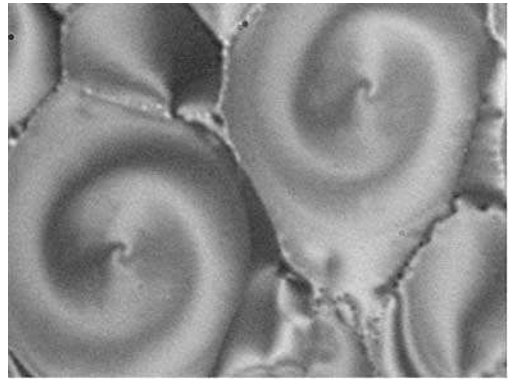,height=5cm}}\vspace{0.25cm}
\caption{ Reversing-spiral pattern observed in a polar film
between slightly uncrossed (vertical and horizontal) polarizers.
Image width: 156 $\mu$m.} \label{spiral}
\end{figure}

The samples were   examined  both under a  polarizing transmission
microscope (PTM) to study  the texture of the $c$-director, and
under a white  light reflection microscope to observe  their
thickness.  The  images were recorded by a CCD camera and image
processing hardware to amplify small differences of  intensity.
With  crossed polarizers, the film image in the PTM is marked by
dark brushes  when $\hat{c}(x,y)$ is parallel to either polarizer,
and  by bright brushes when $\hat{c}$ is at  an intermediate
orientation.   Thus,  a  cross-shaped four-arm extinction pattern
corresponds to   a 45 degree rotation of $\hat{c}$ between  dark
and bright areas   \cite{muzny}. However, because we slightly
uncrossed the polarizers   in order to increase  the light
intensity, the dark/bright   sequence  in Fig.\ \ref{boojum} and
Fig.\ \ref{spiral} corresponds to  a 90 degree rotation  of
$\hat{c}$.

Many interesting effects can occur in thin film samples, including
surface induced phase transitions, and superficial ordering
different from that in the interior of the film.  In the
experiments described here, our observation of rapid fluctuations
of the $c$-director in all cases convinced us that samples were in
the Sm~C* phase.  Therefore we analyzed all our observations in
the thin film limit, in which film properties are considered
uniform in the $z$ direction, resulting from an average of
molecular interactions at the surfaces and in the interior of the
film.  In particular, both surfaces of a film are naturally polar.
In a free standing film, because of the top-bottom symmetry, the
opposite polarities of the two surfaces average to zero, while in
our samples this symmetry is broken, resulting in a net polarity
of the film. However, the net polarity arises only from the
surface layers, so the strength of the polar effects is inversely
proportional to the sample thickness, making that a useful
variable for tuning polarity.

%\section{Observation}
Our general observations  are  summarized as follows.     Ordinary
free  standing films   of   CS1015  exhibit a smectic~C* phase for
all film  thicknesses. The liquid contact induces a non-tilted
Smectic~A phase \cite{Degennes} for extremely thin  films, while
thicker films remain in the tilted Sm~C* phase.  Polar films on
water, ethylene glycol and PVA solutions behave  similarly, but on
SDS solutions we found only the smectic~A phase, perhaps due to
the entry of SDS  into the film.

Upon  forming the polar film, we  see  a multitude of edge
dislocation loops fixing  domains of  different thickness
throughout  the area   of the film (Fig.\  \ref{boojum}). These
defects, roughly circular, are practically static. This is
different from what is observed  in   newly  formed  free standing
smectic films \cite{007}, which steadily evolve toward a uniform
film thickness. Moreover, the domains are  not isolated from each
other, but are packed together,  so they resemble  a foam, rather
than free circular regions.  Within a single  domain  we can
compare  the texture of the c-director to that observed in  the
circular islands (thicker regions) seen in free standing films.

The circular topology  of a domain,  together with a  strong
anchoring of the $c$-director at a fixed angle relative to  the
boundary, combine   to  produce  a  topological constraint,
resulting in the existence of a    point defect in the domain,
around   which the $c$-director   rotates  once.  The detailed
texture within the domain is dictated by the free energy for the
$c$-director, including anchoring energies at the outer boundary
and at the point defect, and the curvature elastic energy for
gradients of the $c$-director within the domain. The free energy
expression for the $c$-director field contains terms consistent
with the symmetry of the film. The coefficients of the polar terms
should vary inversely with film thickness, and all the
coefficients should depend on the average tilt angle of the
director $\hat{n}$ in the Sm~C* phase.  Because the texture
minimizing the free energy depends on a competition among
different terms, we expect to see different characteristic
textures for different samples.

Following the detailed symmetry analysis presented by Langer and
Sethna \cite{LangerSethna}, the elastic free energy $F$ for the
$c$-director in a topologically circular domain of a chiral polar
Sm~C* thin film has the form
\begin{eqnarray*}
F & = &\!\!\! \int\limits_{Area}\!\! \!\! dA \! \left \lbrack K_s
(\nabla \cdot \hat{c})^2\! + K_{sb} (\nabla \cdot \hat{c} )(\nabla
\times \hat{c})+ K_b (\nabla  \times \hat{c})^2 \right \rbrack
\\ & & + \oint\limits_{Boundary} \!\!\! dl \left \lbrack \lambda_s
(\hat{c} \cdot \hat{m}) + \lambda_b (\hat{c} \times \hat{m})
\right \rbrack ,
\end{eqnarray*}
in which  $K_s$ and $K_b$   are the familiar splay and  bend
curvature elastic  constants, $\lambda_s$ and $\lambda_b$ give the
strength of the lowest order terms dictating the $c$-director
anchoring at the domain boundary, and $\hat{m}$ is the outward
normal to the boundary. For 2D vectors, the $\times$ operation
produces a scalar.  In addition to the requirement that $K_s$ and
$K_b$ be positive, to make the free energy density positive
definite for arbitrary values of splay and bend, we require that
$K_{sb}^2<4K_sK_b$.  The $K_s$ and $K_b$ terms could have been
written to include spontaneous splay and bend in the ground state
\cite{BobPershan}, but those added terms can be integrated by
parts to combine with the $\lambda_s$ and $\lambda_b$ terms.

For a free standing film, which is not polar, the $K_{sb}$ and
$\lambda_s$ terms are absent from the free energy. The $\lambda_b$
term favors one direction of tangential orientation of the
$c$-director at the boundary of the island.  If $K_b<K_s
$\cite{B79}, then a simple minimum energy texture is one of pure
bend, with azimuthal orientation of $\hat{c}$, within the circle,
and the point defect at its center. This highly symmetric texture,
often seen in islands on free standing films \cite{011}, is
characterized by a cross-shaped four-brush extinction pattern with
straight arms, when viewed between crossed polarizers
\cite{muzny}.

In the case of a polar film,  this simple texture is no longer
observed, and other textures of the same topology are seen, the
boojum texture (Fig.\ \ref{boojum}) with a point defect located at
the edge of the domain and straight brushes emerging radially from
the defect, and  several different spiral textures, including the
``reversing spiral'' (Fig.\ \ref{spiral}) with a centered defect
around which  the $c$-director field winds first one way and then
the other, as a function of radius.  In thinner films, the boojum
texture is the most common, while spirals are seen in thicker
samples; this may reflect differences in the strength of the polar
perturbation. To understand certain aspects of these textures, we
look to the broken polar symmetry. The free energy expression
above for the polar films contains two terms ($K_{sb}$ and
$\lambda_s$) that break symmetry in different ways.

First, the $\lambda_s$ term in the boundary energy combines with
the $\lambda_b$ term to favor one oblique orientation angle
$\phi^*$ of the $c$-director relative to a domain boundary, rather
than tangential or normal orientation. This oblique boundary
condition is clearly seen in the boojum domains of Fig.\
\ref{boojum}, where dark brushes, corresponding to loci of
constant $\hat{c}$, meet the circumference where it is oriented
obliquely, rather than horizontally or vertically.

Second, the $K_{sb}$ term, discussed first by Langer and Sethna
\cite{LangerSethna}, appears only with both chiral and polar
symmetry. This term lowers the symmetry of a texture because it
favors combined splay and bend deformations. For a circular island
with a centered point defect at $r=0$, instead of pure splay or
pure bend being the lowest energy distortion, a particular ratio
of splay to bend minimizes the energy. In the absence of boundary
anchoring conditions, this would lead to a simple spiral texture,
with the $c$-director making a constant angle $\phi=\phi_{sb}$ (or
$\phi_{sb}+\pi$), with respect to the radial vector throughout the
domain.

Combining these two symmetry breaking effects, it is clear why
various spiral patterns are observed when the point defect is
centered in a domain of a polar chiral film. At the outer boundary
of the domain, the $c$-director is anchored at some angle
$\phi^*=\phi_b$.  At the inner boundary defined by the point
defect, the anchoring angle is $\phi^*=\phi_d$, which need not be
the same as $\phi_b$. Simultaneously, away from the boundaries,
$\phi$ would tend to approach $\phi_{sb}$. The energy minimizing
texture requires that $\phi$ is a function of radius $r$ in the
domain. To find this functional dependence, one can make a
transformation $\xi=\ln(r/\epsilon)$, where $\epsilon$ is the core
radius of the defect.  The problem then maps onto that of a strip
of Sm~C* in the $(\xi,\eta)$-plane contained between two parallel
lines at which the boundary conditions on $\phi$ are $\phi_d$ and
$\phi_b$ respectively, with an applied uniform magnetic field at
angle $\phi_{sb}$ relative to the normal to the strip.  The
director tends to lie parallel to the magnetic field, and the
strength of the effective field coupling depends on relative
magnitudes of the elastic constants. Often, this leads to
solutions in which $\phi$ varies monotonically across the strip,
corresponding to ordinary spirals for the brushes observed in our
experiments. However, much as in the well-studied Fredericks
transition in nematic layers \cite{Degennes}, the solution can
involve $\phi$ first rotating in one direction (from $\phi_d$
toward $\phi_{sb}$)as a function of $\xi$, reaching an extremum,
and then rotating in the other direction to match the second
boundary condition, $\phi_b$. This produces the reversing spiral
observed in Fig.\ \ref{spiral}.

The straight brushes emerging from the defect in the boojum
texture (Fig.\ \ref{boojum}) are in contrast to the spiral
patterns described above. Clearly the $c$-director does not have a
constant orientation angle $\phi_d$ at the inner boundary around
the point defect.  Ignoring anchoring effects at that inner
boundary, one can find an energy minimizing solution which
produces the observed straight brushes and constant $\phi_b$ at
the outer boundary.   In terms of polar coordinates $(r,\theta)$
centered on the point defect, and $\phi$ again an angle relative
to the radial vector, a stable solution is $\phi = \theta +
\phi_b$, independent of $r$. This solution, in a circular domain
of radius $R$ centered at the point $(R,0)$ remarkably has the
fixed angle $\phi_b$ between $\hat{c}$ and $\hat{m}$ at all points
on the boundary of the domain.  This is just the structure we see,
subject to perturbations from the non-circular shape of the
domains.

This boojum is similar  to the topological defect texture observed
in isolated domains of other systems like Langmuir monolayers on
water \cite{sophie}, which  are also polar, and islands of the
Smectic~I phase in an  otherwise Smectic~C free standing film
\cite{LangerSethna}.   In those systems  the anchoring of the
$c$-director  at the  lateral boundary   is weak, and the
topological defect is ``expelled'' from the domain. This lowers
the free  energy by removing the  high elastic strain near the
defect as well as its core energy, and by replacing it with a
lower energy that results from the violation of  the boundary
anchoring in a small region.  However, in our polar films, the
boundary anchoring of the  $c$-director is apparently stronger,
and the defect appears to  be right at the edge of the domain. We
speculate that this strong anchoring results from the large change
in layer thickness at the domain boundaries of our polar films.
Why the defect is attached to the boundary is not known, but it
probably involves the energetics of the defect core structure.

%\section{conclusion}
In conclusion, we have developed  a novel technique for studying
polar smectic  liquid  crystal films, and   have observed  novel
textures in smectic~C* films  in a  polar  environment.  The
strength  of the polar perturbation  can be changed by varying the
film thickness.  We have shown through a free energy analysis that
important aspects of the changes in texture relative to ordinary
free standing smectic~C* films are a direct consequence of the
polar symmetry of the films. In connection with the studies
reported here, we have also observed reversing spirals and other
textural novelties in circular islands on non-polar free standing
Sm~C and Sm~C* films, which will be discussed elsewhere.

%\section{acknowledgment}
We thank Jay  Patel, Darren Link and  Noel Clark  for providing us
with  chemical compounds. I.K. is  grateful  to the city of
Strasbourg for   financial  support  through    a
Strasbourg-Boston  twin cities fellowship.  This research was
supported    by the NSF through   grant DMR-9415656 and   by the
Martin Fisher School   of Physics at Brandeis University.


\begin{thebibliography}{100}
\bibitem{008} This statement   is confirmed by  experiments done
with SCE10 sold by Merck [SmI (-20$^\circ$C) SmC* (61$^\circ$C) N*
(109$^\circ$C) Iso], which has a pitch of 14.6 $\mu$m, 5  times
larger   than CS1015. It exhibits patterns very similar to those
seen in CS1015.
\end{thebibliography}
\end{document}